*Journal of Environmental and Energy Economics*

*Research Article*

# Role of AI Innovation, Clean Energy and Digital Economy towards Net Zero Emission in the United States: An ARDL Approach

Adita Sultana[1], Abdullah Al Abrar Chowdhury[1], Azizul Hakim Rafi[1], Abdulla All Noman[2]

[1]Information Technology of Science, American National University, 1813 East Main Street, Salem, VA 24153.

[2]Montclair State University, Montclair, NJ, USA 07043

Corresponding author: Azizul Hakim Rafi, Email: rafiazizul96@gmail.com,

## Abstract

The current paper investigates the influences of AI innovation, GDP growth, renewable energy utilization, the digital economy, and industrialization on $CO_2$ emissions in the USA from 1990 to 2022, incorporating the ARDL methodology. The outcomes observe that AI innovation, renewable energy usage, and the digital economy reduce $CO_2$ emissions, while GDP expansion and industrialization intensify ecosystem damage. Unit root tests (ADF, PP, and DF-GLS) reveal heterogeneous integration levels amongst components, ensuring robustness in the ARDL analysis. Complementary methods (FMOLS, DOLS, and CCR) validate the results, enhancing their reliability. Pairwise Granger causality assessments identify strong unidirectional connections within $CO_2$ emissions and AI innovation, as well as the digital economy, underscoring their significant roles in ecological sustainability. This research highlights the requirement for strategic actions to nurture equitable growth, including advancements in AI technology, green energy adoption, and environmentally conscious industrial development, to improve environmental quality in the United States.

## Introduction

The importance to ecological conservation and sustainability has attained unparalleled prominence [1,2]. Carbon dioxide ($CO_2$) is a prevalent greenhouse gas (GHG) that retains surface heat in the atmosphere, inhibiting its escape into space and contributing to the rise in global temperatures [3,4,5]. This has prompted countries and international organizations to seek global solutions for reducing carbon emissions and addressing climate change [6,7] . Recently, the USA, the second-largest producer of GHGs in 2017, has a target to reduce GHG emissions by around 27% by 2025, relative to 2005 emission levels [8].The US was selected based on several compelling factors. People regard the United States as the leading entity in energy usage. From 1980 to 2019, quantities of carbon monoxide, lead, and sulfur dioxide decreased by almost 80 percent; the particles had comparable reductions, although the surface levels of ozone diminished by about one-third [9,10]. The country uses the highest amount of renewable and non-renewable resources for each resident [11,12]. In 2020 the United States released emissions totaling 5,416 metric tonnes of $CO_2$ which constituted approximately 16% of worldwide emissions [13]. AI research among nations ranks both China and United States as major global forces in AI development [14]. The USA faces consider obligation in the ecological emergencies and rising temperatures because it ranks as one of the top greenhouse gas manufacturers worldwide thus assessing its environmental sustainability stands as a critical matter. It is essential to recognize the importance of economic growth, energy consumption, industrialization, AI innovation, and the digital economy, especially for a developing nation like the USA. The research drives from global climate crisis demands to reduce $CO_2$ emissions yet maintain economic development and advancing technology. As one of the top economies worldwide the United States requires methods to support industrial growth along with digital expansion while achieving lower ecological influence. This study investigates $CO_2$ emission effects from AI innovation together with digital economy power and renewable energy and traditional factors of GDP growth and industrial development. The relationships between these factors remain essential to create policies which support sustainable economic progress. The research bases its analysis on the ARDL framework by using tests for unit roots along with causality assessments for confirming the findings. This research provides policymakers with essential knowledge about AI innovations along with digital economies as emission reduction agents before examining the detrimental environmental effects of uncontrolled economic expansion and industrialization. These results demonstrate the necessity of directing financial resources towards sustainable AI applications and green energy transformation and environmentally friendly industrial expansion. The research findings contribute to environmental economic scholarship while giving the United States specific policies to reach sustainable ecological and economic harmony.

Studies indicate that both environmental resource depletion and the climate change crisis have strengthened the immediate need for energy efficiency advancements [15]. As the world's primary contributor, the United States assumes a fundamental part in bringing global carbon emissions to zero by 2050 by working to solve the upcoming climate emergency during the present and future times [16,17]. Habitable planet preservation relies on advancing social development and economic conditions together with renewable energy source expansion [18,19]. The Biden administration presents a $2 trillion plan for renewable energy development while enhancing infrastructure and conducting various climate-dependent projects to achieve net-zero emissions by 2050 [20]. Furthermore, the United States offers substantial fossil fuel subsidies, totaling approximately 0.6 trillion USD, making it the second-highest globally. To attain carbon neutrality by diminishing fossil fuel consumption, it is imperative to curtail policy support [21]. Moreover, the United States must promptly reduce its $CO_2$ emissions by employing carbon capture technologies at power and industrial sites, in conjunction with geological storage solutions [22,23].

Theoretically, the link within the digitalized economy and environmental quality is complex and multifaceted. Digital transformation enhances ICT utilization; thereby imposing greater environmental strain through increased energy consumption is related to the manufacturing, and usage of ICT-related items [24]. Enhancing economic growth is the foremost priority for many nations, particularly developing and developing region, to improve their citizen's quality of life [25]. Despite the benefits, heightened industrial productivity is the primary contributor to trash generation and energy consumption [26]. In 2020, the digital economy of the USA amounted to US$13.6 trillion [27]. The prevailing perspective posits that economic expansion adversely impacts the environment during the initial phases of development and then benefits it in later stages [28,29,30]. Public opinion toward these issues parallels the way people respond to climate change since the observable pattern represents an unavoidable trend [31]. Throughout the past 50 years, academic scientists have resurfaced their interest in AI research [32]. It can be anticipated that artificial intelligence will significantly impact global environmental outcomes, productivity, inclusivity, and equity. The influence of AI on sustainable development has been ambiguous. Artificial intelligence (AI) has developed as a formidable instrument across various industries and presents significant potential for government, society, and the economy [33].

The primary aim is to analyze the implication of GDP, the digital economy, clean power usage, industrialization and AI innovation on $CO_2$ emission levels in the USA from 1990 to 2022. After a comprehensive examination of current academic literature, we assert the innovative nature of this research study, substantiated by several foundational ideas. This research presents three notable contributions: Currently, academic research has not specifically analyzed the consequences of green energy utilization, the digital economy, and AI innovation on $CO_2$ emissions in the USA, despite the country's crucial function in the climate change agenda. The USA merits specific scrutiny in the analysis owing to its position as a highly industrialized country that utilizes substantial natural resources, hence imposing considerable environmental strain. Consequently, the USA ranks as the second-highest polluter of $CO_2$ [34]. The data indicates that the USA possesses considerable potential to enhance its energy portfolio by using renewable energy sources. Therefore, the USA's carefully designed and judicious nuclear energy policy could successfully mitigate its air pollution challenges in the near future. Furthermore, we scrutinize the link within industrialization and biodiversity condition. Unlike previous researches, this analysis employs a newly developed econometric technique known as ARDL simulation. The extended IPAT models, which incorporate renewable energy, AI innovations, and a digitalized economy, employ this technique. This methodology obtains, activates, and autonomously produces charts that show misleading changes in the endogenous factor based on the exogenous variable, all while accounting for other elements.

In the second part, we look at research that has been done on certain factors by looking at methodology, theoretical frameworks, the building of empirical models, and the estimation methods that were used. A thorough breakdown of the model results appears in "Results and Discussion," and the final segment covers the analysis together with the proposed action.

## Literature Review

Numerous empirical studies investigate how the three factors of GDP growth, industrialization, and sustainable power usage affect $CO_2$ emission levels. Various studies have analyzed the ARDL model, yet most of them examine how GDP growth, together with urbanization and renewable energy consumption, drives environmental outcomes. Most individuals fail to notice how AI technology, together with the digital economy, affects environmental contaminations. There is limited previous research on ecological deterioration in the United States because this field is new to scientific investigation. Previous research enabled the inquiry to choose variables and methodologies while conducting its study. The subsequent part of this work examines multiple query points.

### GDP and $CO_2$

Many research activities concentrate on the link between growing GDP and biodiversity health. Many experts hold the position that GDP growth usually leads to higher $CO_2$ emission levels. The analysis of environmental quality using $CO_2$ emissions records complicates the current situation. The study of economic growth's (GDP) effect on China's environmental sustainability uses ARDL methodology according to Raihan et al. [35]. Data collected from multiple sources show that the expansion of national economies causes strong increases in $CO^2$ emissions. Sheraz et al. [36] conducted research on G20 carbon dioxide emission responses to GDP variables from 1986 to 2018. Results obtained through the FE-OLS method demonstrated that GDP led to an upsurge in $CO_2$ releases throughout the examined period. Koengkan et al. [37] established through their research that economic progress leads to deterioration of environmental quality. Aslam et al. [38] presents an investigation of industrialization as well as its impact on $CO_2$ emissions when coupled with GDP growth. The research evaluates the environmental Kuznets curve by establishing that per capita GDP drives $CO_2$ emissions increases over time  Findings by Raihan et al. [39] and Sikder et al.[40] together with Abbasi et al.[41] and Magazzino et al.[42] showed parallel outcome.  Using ARDL modeling throughout 1977 to 2016 Solarin et al. [43] discovered that Nigerian economic growth inflicts damage on the environment in the beginning but eventually shows favorable consequences. Mohsin et al. [44] studied the ecological and economic relationship in European and Central Asian regions. Analysis through ARDL technique demonstrated that $CO_2$ emissions and GDP show an opposite sustained link and a positive short-term connection thus indicating GDP expansion damages the ecosystem.

### AI innovation and CO₂

The significant impact of AI technologies, such as machine learning (ML), deep learning (DL), and big data, can improve environmental quality by decreasing pollutant levels [45]. The global economy faces huge strain from ecological issues and the need for more durable infrastructure, making the integration of AI, ML, and DL in manufacturing operations a crucial component of a comprehensive strategy to adhere to long-term sustainability goals [46, 47]. As research progresses, certain researchers have identified the possible environmental ramifications of AI [48,49,50]. Vinuesa et al. [51] examined the effects of AI on the 17 goals and 169 specific targets specified in the UN's "2030 Agenda for Sustainable Development," demonstrating that AI can aid in the realization of the majority of these targets. Dhar [52] examined the dual function of AI in $CO_2$ emission reduction, emphasizing its position as both a means to combat global warming and a notable source of carbon emissions. Moreover, Chen et al. [53] found that the impact of AI innovation on reducing $CO_2$ emissions is more pronounced in large cities, major urban areas, well-developed infrastructure, and technologically advanced cities, based on 270 Chinese cities.

### Renewable Energy use and CO₂

The long-term cost benefits of alternative or green energy sources will enhance the general standard of living [54]. Environmental disasters arose from increasing fossil fuel usage so green power needs to replace them to protect ecosystems and obtain secure reliable power [55]. Energy efficiency methods work for ensuring green ecosystem and develop equitable growth through the deployment of sustainable and clean energy resources [56]. Baloch et al. [57] analyzed the connection within renewable energies and $CO_2$ emissions in BRICS countries by using the AMG estimator from 1990 to 2015. Research findings demonstrated that clean energy caused decreased $CO_2$ emissions throughout the entire BRICS coalition except for South Africa. Dogan and Ozturk [58] study the implication of renewable and non-renewable energy utilization on $CO_2$ emissions throughout the 1980-2014 periods in the United States. Research findings demonstrate that raising renewable energy usage creates effective reduction of environmental harm throughout extended periods. A research conducted by Salahuddin et al.[59] focused on SSA countries and Kartal et al.[60] worked with USA while Dagar et al. [61] did their study with the OECD economies all reaching similar findings. Numerous experimental studies show renewable energy implementation produces minor effects on $CO_2$ emissions while producing possible adverse environmental consequences from increased GHG output. The research by Apergis and Payne [62] demonstrates that renewable energy technology failed to reduce emissions within a short-term period across 19 developing economies and industrial nations. In the short term Farhani [63] finds that REN output creates a causal link to $CO_2$ emissions yet this effect disappears in the long term. Several studies demonstrate renewable energy utilization brings adverse effects to ecosystems according to Murshed et al.[64] in G-7 countries as well as Abbas et al. [65] in BRICS region and Silva et al.[66] in Africa.

### Digital Economy and CO₂

The modernization of the nature hypothesis demonstrates how digital technological advancement offers solutions to ecological problems while producing theoretical analysis for digital economy-based sustainability [67]. The paper of Wang et al. [68] introduces a multifaceted digital economy index that tracks Chinese provincial data from 2006 to 2017 while studying digital commerce connections with levels of $CO_2$ emissions. Through their analysis, which used system-GMM, they found evidence that DE operations create negative effects on $CO_2$ emissions. Ma et al. [69] examined China's digital economy capability for minimizing pollutants. The study indicates that the DE of China plays a substantial role in reducing $CO_2$ emissions. The digital economy emerges as both a promising and innovative environmental sustainability strategy because of rising environmental awareness and a growing economic need for sustainable solutions [70]. Shobande and Ogbeifun [71] replicated the findings along with Kovacikova et al. [72] in their own regional studies. Xu et al. [73] explored the current and quantitative patterns that relate the DE to natural world throughout China's 287 prefecture-level cities from 2008 to 2018. Research outcomes demonstrate that the relationship between the DE and ecological damage runs in reverse directions as it displays complex spatial and temporal patterns. According to Nguyen et al. [74], the increasing scale of economic activities facilitated by the DE increases environmental emissions by using more energy. Research by Li and Wang [75] established that the correlation between the DE and $CO_2$ releases followed an inverted U pattern.

### *Industrialization* and $CO_2$

The detailed linkage between industrial development and biodiversity reduction became a major topic when nations aim to expand their economic base without causing environmental deterioration. In their study Sumaira and Siddique [76] investigated how industrialization creates cause and effect patterns with $CO_2$ emissions which operate in both directions. The investigations took place in the SAARC area from 1984 to 2016. Through their analysis Ahmed et al. [77] evaluated how industrialization affects environmental condition in the Asia-Pacific region. The results using ARDL methods showed that industrial growth has a major beneficial impact on environmental conditions. Opoku and Aluko [78] conducted research that analyzed various environmental results of industrialization from 2000 to 2016 across 37 African nations. The researchers established that national industrial growth decreases environmental destruction. Sikder et al. [40] investigated emissions based on industrial development among 23 developing nations during the period from 1995 to 2018. According to the ARDL model industrialization grows by 0.54% when $CO_2$ emissions increase by 1% throughout the long-term period. Research by Nasir et al. [79] investigates ecological damage elements in Australia throughout the 1980-2014 periods. Their analysis using EKC and STIRPAT adopted complete framework to demonstrate that industrial development shows no significant link with $CO_2$ emissions. Mentel et al.[80] researched Africa while Kermani et al. [81] investigated Iran, Xu and Lin [82] studied China and Farooq et al.[83] analyzed India and all these studies demonstrated that $CO_2$ emissions rise because of industrial growth and harm environmental health.

### *Literature Gap*

The research addresses critical information deficiencies by examining the United States and its distinct macroeconomic and environmental attributes. Despite the global focus on equitable growth, there have been limited, thorough studies conducted in the USA that investigate the cumulative implication of industrialization, AI innovation, and the digital economy on carbon intensity. Comprehensive assessments of the complex interrelations among these factors are frequently absent in the current literature, especially with the ARDL framework. Conflicting findings continue on the relationship among DGE, AI innovation, and $CO_2$ emissions, despite earlier research acknowledging the necessity for more thorough examinations of these associations. It is essential to recognize that technical growth, along with the adoption of alternative energy and a digital economy, can promote the utilization of cutting-edge, environmentally friendly technologies, thereby facilitating a sustainable world. Therefore, the objective of the article is to address these inadequacies and provide policymakers with essential data to develop sustainable plans for decarbonizing emissions.

## Methodology
### *Data and Variables*

This analysis utilized time series data from 1990 to 2020 for the United States. The World Development Indicator (WDI) supplied the data on $CO_2$ emissions, which is employed as an endogenous variable. Likewise, information regarding AI innovation and the digital economy is sourced from Our World in Data. Furthermore, the statistics on GDP, renewable energy utilization, and industrialization are sourced from WDI. Table 1 distinguishes the factors, their logarithmic forms, units of measurement, and sources of data.

### *Theoretical framework*
Dietz and Rosa [84,85] developed a revised "Stochastic Impacts by Regression on Population, Affluence, and Technology" (STIRPAT) model to resolve problems with IPAT format [86]. The method predicts irregular functional connections between variables which affect the ecosystem environment [87]. The STIRPAT model enables researchers to add extra independent factors like energy usage when tracking environmental influences [88]. The basic model structure appears in Equation (1).

$$I \equiv P.A.T \ldots\ldots\ldots\ldots\ldots\ldots\ldots\ldots (1)$$

In this case, "P" denotes population number, "A" denotes wealth, "T" denotes advances in technology, and "I" denotes an ecological impact. Researcher examines the following formulation:

$$I_{it} = C P_{it}^{\gamma_1} A_{it}^{\gamma_2} T_{it}^{\gamma_3} \varepsilon_{it} \ldots\ldots\ldots\ldots\ldots\ldots\ldots\ldots\ldots (2)$$

Table 1. Data and variables

| Variables | Description | Logarithmic Form | Unit of Measurement | Source |
|---|---|---|---|---|
| $CO_2$ | $CO_2$ Emission | $LCO_2$ | $CO_2$ Emission (kt) | WDI |
| GDP | Gross Domestic Product | LGDP | GDP per capita (current US$) | WDI |
| AI | AI Innovation | LPAI | Estimated Investment in AI (US$) | Our World in Data |
| REN | Renewable Energy Use | LREN | Renewable Energy Use (% of total energy use) | WDI |
| DGE | Digital Economy | LDGE | ICT good imports (% of total goods imports) | Our World in Data |
| INDUS | Industrialization | LINDUS | Industry (including construction), value added (current US$) | WDI |

The logarithmic transformation can stabilize data, compress variable scales, and reduce model heteroscedasticity and collinearity while maintaining data structure and correlation. Understanding unit differences in factors affecting carbon intensity is crucial for the research topic. Equation (3) presents the logarithmic representation.

$$LnI_{it} = C + \gamma_1 LnP_{it} + \gamma_2 LnA_{it} + \gamma_3 LnT_{it} + \varepsilon_{it} \dots\dots\dots\dots (3)$$

In this context, P symbolizes the population of a nation, A its affluence, and T its technology at time t. The random error component is denoted by $\varepsilon$, while the constant component in the STIRPAT methodology is C. Eqn.(4) is the mathematical framework for this study:

$$CO_{2it} = f(GDP_{it}, AI_{it}, REN_{it}, DGE_{it}, INDUS_{it}) \dots\dots\dots\dots (4)$$

The explanatory variables in this instance are GDP, AI innovation, renewable energy use, digital economy and industrialization whereas the dependent variable is $CO_2$ emission. An alternate way to describe the empirical model in logarithmic form is as follows:

$$LnCO_{2it} = \beta_0 + \beta_1 LnGDP_{it} + \beta_2 LnAI_{it} + \beta_3 LnREN_{it} + \beta_4 LnDGE_{it} + \beta_5 LnINDUS_{it} + \varepsilon_{it} \dots\dots\dots\dots (5)$$

Here, $\beta_0$ to $\beta_5$ is used as the coefficient of five different selected independent variables.

### Empirical Framework
The fundamental purpose of this inquiry analyzes the connection among AI innovation and GDP growth, industrialization and digital economy and energy usage on $CO_2$ emissions in the USA region. The research study will execute the following sequence of steps towards its objective. Tests including ADF and P-P and DF-GLS were used to perform unit root examinations. We utilize ARDL modeling to discover the connection patterns in the variables both in the short and long term. The robustness tests were done with various techniques which included FMOLS and DOLS and CCR. Multiple diagnostic assessments were also applied to confirm that the model contained no disturbing factors.

### Unit Root Test
It is imprudent to check the stability of the data prior to examining any correlations between the eras [89]. Whether the dataset exhibits stationarity in integrated order zero (I(0)) or integrated order one (I(1)), the current study first examines the links between the response and its independent factors. The evasion of the I(2) sequence is considered invalid and may lead to erroneous results [90]. This study employs the ADF test [91], DF-GLS test [92] and PP test [93] to assess the stability of the variables.

### ARDL Structure
Pesaran et al. [94] introduced the ARDL limits analysis as a cointegration method that we used for assessing the lasting associations between factors. The cointegration test provides superior sequencing of integration when compared to traditional methods. This analytical method applies when parameters demonstrate I(1) and I(0) stability or an I(1)/I(0) combination status [95]. The ARDL framework determines cointegration through the ARDL F-statistic which computes its results using variable lag structures optimized for individual variables [96]. Cointegration between the parameters becomes evident when the values of the ARDL F-statistic exceed the predefined upper threshold. The absence of cointegration exists among the variables when the ARDL F-statistic falls beneath the lower critical boundary [97]. Eq. (6) applies the ARDL bound analysis to determine cointegration.

$$\Delta LCO_{2t} = \tau_0 + \tau_1 LCO_{2t-1} + \tau_2 LGDP_{t-1} + \tau_3 LAI_{t-1} + \tau_4 LREN_{t-1} + \tau_5 LDGE_{t-1}$$
$$+ \tau_6 LINDUS_{t-1} + \sum_{i=1}^{q} \gamma_1 \, \Delta LCO_{2t-i} + \sum_{i=1}^{q} \gamma_2 \, \Delta LGDP_{t-i} + \sum_{i=1}^{q} \gamma_3 \, \Delta LAI_{t-i}$$
$$+ \sum_{i=1}^{q} \gamma_4 \, \Delta LREN_{t-i} + \sum_{i=1}^{q} \gamma_5 \, \Delta LDGE_{t-i} + \sum_{i=1}^{q} \gamma_6 \, \Delta LINDUS_{t-i} + \varepsilon_t$$

$$(6)$$

where $\Delta$ is the first difference operator, and q indicates the length of the lag that is optimal.

The ECM method produces consistent outcome even with comparatively small samples [98]. The ECM amalgamates short-term nuances with long-term stability to maintain a comprehensive perspective [99]. The symbol $\theta$ represents the coefficient of ECM. Equation (7) is used to explore short run associations of the variables.

$$\Delta LCO_{2t} = \tau_0 + \tau_1 LCO_{2t-1} + \tau_2 LGDP_{t-1} + \tau_3 LAI_{t-1} + \tau_4 LREN_{t-1} + \tau_5 LDGE_{t-1}$$
$$+ \tau_6 LINDUS_{t-1} + \sum_{i=1}^{q} \gamma_1 \, \Delta LCO_{2t-i} + \sum_{i=1}^{q} \gamma_2 \, \Delta LGDP_{t-i} + \sum_{i=1}^{q} \gamma_3 \, \Delta LAI_{t-i}$$
$$+ \sum_{i=1}^{q} \gamma_4 \, \Delta LREN_{t-i} + \sum_{i=1}^{q} \gamma_5 \, \Delta LDGE_{t-i} + \sum_{i=1}^{q} \gamma_6 \, \Delta LINDUS_{t-i}$$
$$+ \theta ECM_{t-1} + \varepsilon_t$$

$$(7)$$

***Robustness Check***
The study evaluated ARDL results through alternative cointegration regression methods that included FMOLS by Hansen and Phillips [100] as well as DOLS methodology by Stock and Watson [101] and CCR test by Park [102]. The adoption of these methods emerged because of two main requirements [103]. The I(1) parameters need to exhibit cointegration before implementing any of FMOLS, DOLS, or CCR methods. The application of these methods produces consistent parameters while using small sample sizes in testing. The methods address endogeneity and serial correlation and omitted variable bias and measurement errors of parameters. The results produced by these methods become more and more efficient as the sample size increases [104].

***Pairwise Granger Causality Test***
The research utilized Granger-causality test developed by Granger [105] to verify connections among its components. The concept stands as a predictive statistical procedure that brings multiple benefits compared to other approaches when working with time series data [106]. This test provides the crucial benefit of simultaneous analysis of multiple lags by reducing the impact of elevated lag orders [107]. A time series Y demonstrates "Granger-causality" to another time series X through its ability to enhance future prediction of X values. At time t the time series values for both variables are denoted by Xt and Yt. The bivariate autoregressive model successfully demonstrates how the variables X and Y operate.

$$X_t = \beta_1 + \sum_{i=1}^{n} \alpha_i Y_{t-i} + \sum_{i=1}^{n} \mu_i X_{t-1} + e_t$$

(8)

$$Y_t = \beta_2 + \sum_{i=1}^{n} \Omega_i Y_{t-1} + \sum_{i=1}^{n} \infty_i X_{t-i} + u_t$$

(9)

### *Diagnostic Tests*

The research employed diagnostic assessments such as the Jarque-Bera test, Lagrange Multiplier test, and Breusch-Pagan-Godfrey test to verify model assumptions and guarantee robust outcomes. The Jarque-Bera test evaluates the normality of residuals, whereas the Lagrange Multiplier test identifies serial correlation within residuals. The Breusch-Pagan-Godfrey test assesses heteroscedasticity, which may result in erroneous estimates and standard errors. Mitigating heteroscedasticity enhances model precision and inference dependability.

## Results and Discussion

### *Summary Statistics*

Table 1 contains descriptive statistics that have been presented as a summary. The evaluation and analysis of collected data reveal identical median and mean values across all variables. The distribution of all variables remains normal because their skewness approaches zero points and kurtosis stays below 3 while their Jarque-Bera test statistics fall under their thresholds.

Table 02: Summary statistics.

| Variable | Obs | Mean | Std. Dev. | Min | Max |
|----------|-----|------|-----------|-----|-----|
| T | 34 | 2003.676 | 11.726 | 1990 | 2021 |
| $LCO_2$ | 34 | 10.376 | .732 | 9.29 | 11.472 |
| LGDP | 34 | 6.402 | .72 | 5.145 | 7.807 |
| LPOS | 34 | -.985 | .428 | -1.864 | -.371 |
| LEDU | 34 | .627 | .123 | .355 | .798 |
| LFDI | 34 | 19.031 | 2.589 | 14.145 | 21.764 |
| LPOP | 34 | 18.708 | .2 | 18.105 | 18.948 |

### *Unit Root test*

The analysis shows the stationarity findings of unit root test using both I(0) and I(1) first-difference forms in Table 03. It indicates that industrialization serves as the sole variable showing stationarity at level I(0) while $CO_2$, GDP, AI innovation, green power utilization and digital economy exist in non-stationary form before subtracting the first differences. The differently integrated series require us to initiate assessment then proceed with applying the ARDL modeling framework.

Table 3. Results of unit root test.

| Variables | ADF | | P-P | | DF-GLS | | Decision |
|-----------|-----|-----|-----|-----|--------|-----|----------|
| | I(0) | I(1) | I(0) | I(1) | I(0) | I(1) | |
| $LCO_2$ | -0.644 | -4.647*** | -0.625 | -4.014*** | -0.427 | -4.302*** | I(1) |
| LGDP | -0.732 | -4.271*** | -0.740 | -4.739*** | -0.761 | -3.051** | I(1) |
| LAI | -0.502 | -4.615*** | -0.597 | -4.523*** | -0.872 | -3.365** | I(1) |
| LREN | -1.321 | -4.321*** | -1.034 | -4.320*** | -1.431 | -4.623*** | I(1) |
| LDGE | -1.025 | -5.713*** | -1.011 | -5.166*** | -0.806 | -4.453*** | I(1) |
| LINDUS | -4.340*** | -5.787*** | -4.120*** | -4.243*** | -4.981*** | -5.462*** | I(0) |

Table 4. Results of ARDL bound test.

| | Test Statistics | Value | K | |
|---|---|---|---|---|
| | F statistics | 5.0348 | 5 | |
| | Significance level | | | |
| Critical Bounds | 10% | 5% | 2.50% | 1% |
| I(0) | 1.98 | 2.29 | 2.60 | 2.98 |
| I(1) | 3.01 | 3.24 | 3.71 | 3.99 |

### ARDL bound test

Following the verification of the variable's unit roots, this investigation employed the ARDL bounds test to examine the nature of the long-term relationship between the variables. Table 4 presents the empirical findings derived from the ARDL-limits testing methodologies for cointegration. The calculated F-statistic (5.10348) was higher than the upper critical bound values. This means that there was long-term cointegration within the selected factors.

### ARDL result

Table 5 utilizes the ARDL model to examine short-term and long-term influences of LGDP, LAI, LREN, LDGE, LINDUS on $LCO_2$ found within the United States. A 1% boost in LGDP leads to a rise of 0.332% in $LCO_2$ levels during the long-term period along with a short-term impact of 0.142%. The findings demonstrate that rising GDP levels produce increased $CO_2$ emissions because financial growth brings about more manufacturing operations and energy consumption and asset deployment. Several researchers have shown that economic expansions through increased GDP production create negative environmental consequences. Research evidence supporting this connection can be found in publications by Voumik and Sultana [108] Majeed et al. [109] Kirikkaleli et al.[110] and Qayyum et al.[111]. Research from Zubair et al.[112], Ali et al.[113], Halliru et al.[114] opposed the positive connection within GDP and $CO_2$ emission. A 1% boost in LAI causes to a decline of $LCO_2$ in both time scenarios by 0.115% and 0.076%. The use of AI technologies in the United States generates substantial environmental sustainability benefits based on these research findings. Thus the research from Cifuentes et al.[115], Ridwan et al.[116], Ham et al.[117] alongside Chattopadhyay et al.[118] shows that customized AI techniques must be developed to foster global sustainability goals. Moreover, Awan et al.[119] shows innovation leads to increased pollution while recommending the adoption of pollution-minimized technologies. The analysis shows that LREN positively affects $LCO_2$ in both time periods with confirmed statistical importance. The USA ecosystem benefits from renewable energy consumption according to the research findings. The relation between LREN and $LCO_2$ demonstrates that $LCO_2$ decreases by 0.193% in the long term and by 0.102% in the short term when there is a 1% increase in LREN. Green energy offers a substitute for fossil fuels through environmentally friendly sustainable power technologies that produce negligible greenhouse gas emissions. The research findings of Sharmin [120] support the findings alongside those of Waheed et al.[121] and Sharif et al.[122]. According to Silva et al.[66] and Yurtkuran [123] and Lee [124] researchers reported negative correlations between clean energy usage and pollutant level in African countries as well as Turkey and European economies respectively.

At the same way, there is a favorable link within LDGE and $LCO_2$, with each 1% increment in DGE mitigates the $CO_2$ emission by 0.057% over time and 0.061% immediately. It is significant 1% level and indicating that DGE is beneficially for the ecosystem of the USA. The digital economy leads to reduce $CO_2$ emissions by fostering energy-efficient innovations and decreasing the reliance of conventional industrial operations. Shahbaz et al.[125] and Song et al.[126] corroborated with this conclusion . However, Kuntsman and Rattle [127] assert that digital devices have inflicted significant harm on the ecosystem throughout the manufacturing, preservation, and disposal processes. Similarly, Guo and Liang [128] and Ozturk and Ullah [129] observed same conclusions. Statistical analysis shows that both time frames LURBA growth negatively affects environmental quality according to the LINDUS coefficients. Additional levels of LINDUS result in a 0.7182% increase in LCO2 within the long-term period while generating a 0.204% rise in the short term. Due to higher energy use and reliance on petroleum and natural gas in manufacturing operations, modernization most likely results in more $CO_2$ emissions. Multiple research studies including Sikder et al.[40], Mahmood et al.[130], Khan et al.[131] illuminated how industrialization harmed natural world health. Industrialization achieves environmental sustainability through decreased $CO_2$ releases according to studies performed by Zafar et al.[132], Pong et al.[133] and Elfaki et al.[134].

Table 5. Results of ARDL short-run and Long-run.

| VARIABLES | LR | SR |
|---|---|---|
| LGDP | 0.332***(0.4321) | |
| LAI | -0.115***(0.0313) | |
| LREN | -0.193***(0.3412) | |
| LDGE | -0.057**(0.5431) | |
| LINDYS | 0.182***(0.1337) | |
| D.LGDP | | 0.142**(0.4534) |
| D.LAI | | -0.462***(0.0074) |
| D.LREN | | -0.102***(0.1540) |
| D.LDGE | | -0.061***(0.0074) |
| LINDUS | | 0.204***(0.5464) |
| ECT (Speed Adjustment) | | -0.243***(0.6512) |
| Constant | | 10.910***(11.2423) |
| R-square | 0.8950 | |

***Robustness Check***

The results of the ARDL test are confirmed by three other methods, shown in Table 6: DOLS, FMOLS, and CCR. The coefficients from the FMOLS analysis show statistical significance at the 1% level while producing positive values. A single percentage increment of GDP triggers a 0.443% boost in $CO_2$ emission levels. When LAI increases by one percentage point, the USA experiences a decrease of 0.145 percent in $CO_2$ emissions. A 1% increase in LREN together with LDGE can reduce $LCO_2$ by 0.253% and 0.027%, respectively. The relationship between $LCO_2$ and LINDUS shows a positive trend because elevating LINDUS by 1% generates a 0.168% increase in $CO_2$ emissions. The results confirm that both GDP growth and industrial development produce damaging impacts on the natural environment of the USA. The findings match those obtained from both the short-term and long-term ARDL estimations.

The DOLS model indicates that $LCO_2$ increases by 0.315% and 0.671% on average when LGDP and LINDUS rise by 1%. The CO2 emission levels decrease by 0.156% when LAI increases by 1% along with equivalent increases of 0.258% from LREN and 0.068% from LDGE. According to the CCR analysis, LGDP, LINDUS, and LDGE have effects on $LCO_2$ changes that are, on average, 0.136%, 0.045%, and 0.096%. $LCO_2$ decreased by an average of 0.341% and 0.236% per 1% rise in LAI and LREN, respectively, which matched the ARDL results except for the LAI data. This model confirms the significance of all components at the 5% level with an additional 1% level of significance for LGDP, LREN, and LINDUS. The results from all three assessments demonstrate that the ARDL model reaches reliable conclusions about the data patterns.

Table 6.  Results of Robustness Check

| Variables | FMOLS | DOLS | CCR |
|---|---|---|---|
| LCO$_2$ dependent | | | |
| LGDP | 0.443***(0.8623) | 0.315**(0.4352) | 0.136***(0.4526) |
| LAI | -0.145***(0.0562) | -0.156**(0.0731) | -0.3413**(0.0762) |
| LREN | -0.253***(0.1718) | -0.258***(0.5214) | -0.236***(0.1345) |
| LDGE | -0.027**(0.0823) | -0.068*(0.6720) | 0.045**(0.0820) |
| LINDUS | 0.168**(0.2345) | 0.671**(0.4591) | 0.096***(0.8327) |
| C | 10.708**(6.0127) | 11.3101**(8.5372) | 10.294**(8.9783) |
| R-squared | 0.8913 | 0.9041 | 0.8965 |

***Pairwise granger causality test***

Table 7 delineates the outcomes of the causal linkages across diverse determinants. The results of an F-statistic 4.65823 and p-value .0499 indicate no Granger-causal link between LLGDP and LCO$_2$ because the test rejects the null hypothesis at a 5% significance level. The data indicates single-directional cause-effect relationships between LAI and LDGE and LCO$_2$ since their p-values lie below the standard significance threshold. A two-way causal connection exists within LCO$_2$ and INDUS. The analyzed p-values which exceed the significance threshold demonstrate LCO$_2$ has no statistically significant impact on LGDP, LAI or LDGE. For these interactions we lack enough evidence to deny the null hypothesis stating causality does not exist.

Table 7. Causality test.

| Null Hypothesis | Obs | F-Statistic | Prob. |
|---|---|---|---|
| LGDP $\neq$ LCO | 30 | 4.65823 | 0.0499 |
| LCO$_2$ $\neq$ LGDP | | 0.76382 | 0.647 |
| LAI $\neq$ LCO$_2$ | 30 | 3.78341 | 0.0027 |
| LCO$_2$ $\neq$ LAI | | 0.67394 | 0.7692 |
| LREN $\neq$ LCO$_2$ | 30 | 6.67381 | 0.0072 |
| LCO$_2$ $\neq$ LREN | | 0.46839 | 0.0283 |
| LDGE $\neq$ LCO$_2$ | 30 | 3.89923 | 0.0071 |
| LCO$_2$ $\neq$ LDGE | | 0.67892 | 0.1381 |
| LINDUS $\neq$ LCO$_2$ | 30 | 2.78290 | 0.0077 |
| LCO$_2$ $\neq$ LINDUS | | 3.39028 | 0.0154 |

***Diagnostic Test***

The diagnostic assessment results appear in Table 8. The experimental results proved that all diagnostic procedures yielded minimal effectiveness rates during which the null hypothesis maintained its validity. The JB test results show that the residuals follow a normal distribution since the calculated p value stands at 0.2078. Analysis through the LM method shows that the residuals do not show serial correlation since the p-value stands at 0.5698. The BPG test validates that the residuals show no heteroscedasticity because its p-value reaches 0.7830.

Table 8. The findings of diagnostic tests.

| Diagnostic tests | Coefficient | p-value |
|---|---|---|
| Normality test | 0.26531 | 0.2078 |
| Serial Correlation test | 0.78901 | 0.5698 |
| Heterocedasicity test | 1.3245 | 0.7830 |

The structural reliability assessment of residuals at extended and brief intervals uses CUSUM and CUSUM-SQ statistics. The CUSUM-SQ plot graphically displays results within accepted critical limits through its position on the crucial line as shown in figure 01. The tests support the acceptability and coherence of parameters at a 5% significance level.

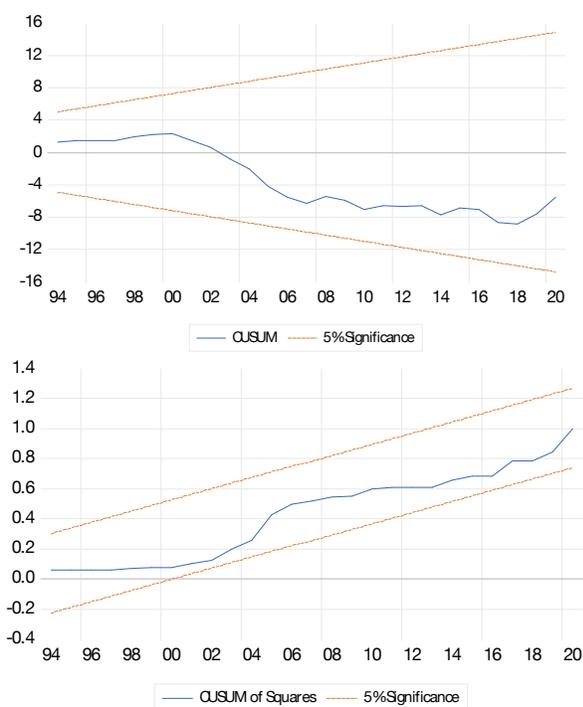

Figure 1. CUMSUM and CUSUM-SQ.

## Conclusion

A thorough examination has studied the implications of AI innovation, GDP growth, cleanenergy usage, digital economy, and industrialization on $CO_2$ emissions in the USA from 1990 to 2022. The research confirms through the ARDL framework that AI innovation, along with the digital economy and renewable energy production, lowers environmental stress, yet GDP growth, together with industrialization, worsens environmental issues. The results from ADF, PP, and DF-GLS tests verify that these variables possess different levels of integration status without existing unit root issues. The analysis through ARDL methodology shows positive relationships among AI innovation and green power adoption and the digital economy toward USA $CO_2$ emissions reduction. The successful deployment of sustainable energy systems and a digitalized economy together with AI technology brings positive results to environmental quality. Economic development, along with industrial evolution, generates an opposing reaction with $CO_2$ emissions, indicating that these activities destructively affect environmental quality. The implementation of energy-efficient approaches and environmentally friendly industrial methods allows the development of innovative rivalries, which lead to sophisticated technology access. The reliability of ARDL results gains additional credibility through robustness assessments that use the combination of FMOLS, DOLS, and CCR. The Granger causality tests reveal that unidirectional causal effects run from $LCO_2$ to LAI and from $LCO_2$ to LDGE. The relationships between economic developments and advancements in AI, together with digitalization, demonstrate extensive effects on environmental sustainability within the USA. The study provides multiple suggested laws to enhance America's sustainable economic growth through technical innovation implementation alongside greener energy consumption methods and sustainable industrial installation.

### Declaration

**Ethics approval/declaration:** Not applicable.
**Consent to participate:** Not applicable.
**Consent for publication:** Not applicable.
**Acknowledgment:** Not applicable.
**Conflict of interest:** The authors declare no conflict of interest.
**Data availability:** Data will be available upon reasonable request from corresponding author.
**Author's contribution:** Adita Sultana and Abdullah Al Abrar Chowdhury were responsible for the conceptualization and design of the study, as well as conducting the literature review. Azizul Hakim Rafi led the development of the methodology and performed the data analysis. Abdulla All Noman contributed to the interpretation of the findings and the drafting of the discussion. All authors contributed to manuscript revision and approved the final version, ensuring the integrity and quality of the work.

## Reference

1. Polcyn, J., Voumik, L. C., Ridwan, M., Ray, S., & Vovk, V. (2023). Evaluating the influences of health expenditure, energy consumption, and environmental pollution on life expectancy in Asia. International Journal of Environmental Research and Public Health, 20(5), 4000. https://doi.org/10.3390/ijerph20054000

2. Voumik, L. C., & Ridwan, M. (2023). Impact of FDI, industrialization, and education on the environment in Argentina: ARDL approach. Heliyon, 9(1). https://doi.org/10.1016/j.heliyon.2023.e12872

3. Pattak, D. C., Tahrim, F., Salehi, M., Voumik, L. C., Akter, S., Ridwan, M., ... & Zimon, G. (2023). The driving factors of Italy's $CO_2$ emissions based on the STIRPAT model: ARDL, FMOLS, DOLS, and CCR approaches. Energies, 16(15), 5845. https://doi.org/10.3390/en16155845

4. Raihan, A., Zimon, G., Ridwan, M., Rahman, M. M., & Salehi, M. (2025). Role of mineral resource rents, renewable energy, and energy efficiency toward carbon neutrality in China. Energy Nexus, 100394. https://doi.org/10.1016/j.nexus.2025.100394

5. Raihan, A., Ridwan, M., Zimon, G., Rahman, J., Tanchangya, T., Bari, A. M., ... & Akter, R. (2025). Dynamic effects of foreign direct investment, globalization, economic growth, and energy consumption on carbon emissions in Mexico: An ARDL approach. Innovation and Green Development, 4(2), 100207. https://doi.org/10.1016/j.igd.2025.100207

6. Ridwan, M., Akther, A., Tamim, M. A., Ridzuan, A. R., Esquivias, M. A., & Wibowo, W. (2024). Environmental health in BIMSTEC: the roles of forestry, urbanization, and financial access using LCC theory, DKSE, and quantile regression. Discover Sustainability, 5(1), 429. https://doi.org/10.1007/s43621-024-00679-4

7. Raihan, A., Atasoy, F. G., Coskun, M. B., Tanchangya, T., Rahman, J., Ridwan, M., ... & Yer, H. (2024). Fintech adoption and sustainable deployment of natural resources: Evidence from mineral management in Brazil. Resources Policy, 99, 105411. https://doi.org/10.1016/j.resourpol.2024.105411

8. Raihan, A., Ibrahim, S., Ridwan, M., Rahman, M. S., Bari, A. M., & Atasoy, F. G. (2025). Role of renewable energy and foreign direct investment toward economic growth in Egypt. Innovation and Green Development, 4(1), 100185. https://doi.org/10.1016/j.igd.2024.100185

9. Ridwan, M., Akther, A., Al Absy, M. S. M., Tahsin, M. S., Ridzuan, A. R., Yagis, O., & Mukhtar, K. J. (2024). The Role of Tourism, Technological Innovation, and Globalization in Driving Energy Demand in Major Tourist Regions. International Journal of Energy Economics and Policy, 14(6), 675-689. https://doi.org/10.32479/ijeep.17344

10. Ridwan, M., Aspy, N. N., Bala, S., Hossain, M. E., Akther, A., Eleais, M., & Esquivias, M. A. (2024). Determinants of environmental sustainability in the United States: analyzing the role of financial development and stock market capitalization using LCC framework. Discover Sustainability, 5(1), 319. https://doi.org/10.1007/s43621-024-00539-1

11. Chowdhury, A. A. A., Rafi, A. H., Sultana, A., & Noman, A. A. (2024). Enhancing green economy with artificial intelligence: Role of energy use and FDI in the United States. arXiv preprint arXiv:2501.14747. https://doi.org/10.48550/arXiv.2501.14747

12. Rafi, A. H., Chowdhury, A. A. A., Sultana, A., & Noman, A. A. (2024). Unveiling the role of artificial intelligence and stock market growth in achieving carbon neutrality in the United States: An ARDL model analysis. arXiv preprint arXiv:2412.16166. https://doi.org/10.48550/arXiv.2412.16166

13. Chowdhury, A. A. A., Sultana, A., Rafi, A. H., & Tariq, M. (2024). AI-driven predictive analytics in orthopedic surgery outcomes. Revista Espanola de Documentacion Cientifica, 19(2), 104-124.

14. Sultana, A., Rafi, A. H., Chowdhury, A. A. A., & Tariq, M. (2023). Leveraging artificial intelligence in neuroimaging for enhanced brain health diagnosis. Revista de Inteligencia Artificial en Medicina, 14(1), 1217-1235.

15. Sultana, A., Rafi, A. H., Chowdhury, A. A. A., & Tariq, M. (2023). AI in neurology: Predictive models for early detection of cognitive decline. Revista Espanola de Documentacion Cientifica, 17(2), 335-349.

16. Rafi, A. H., Chowdhury, A. A. A., Sultana, A., & Tariq, M. (2024). Artificial intelligence for early diagnosis and personalized treatment in gynecology. International Journal of Advanced Engineering Technologies and Innovations, 2(1), 286-306.

17. Tipon Tanchangya, M. R., Raihan, A., Khayruzzaman, M. S. R., Rahman, J., Foisal, M. Z. U., Babla Mohajan, A. P., ... & Islam, S. Nexus Between Financial Development and Renewable Energy Usage in Bangladesh. https://doi.org/10.56946/jeee.v2i1.524

18. Waqar, M., Zada, H., Rafi, A., & Artas, A. (2023). Asymmetry in Oil Price Shocks Effect Economic Policy Uncer-tainty? An Empirical Study from Pakistan. Jinnah Business Review, 11(1).

19. Rahman, J., Foisal, M. Z. U., Mohajan, B., Rafi, A. H., Islam, S., & Paul, A. Nexus Between Agriculture, Industrialization, Imports, and Carbon Emissions in Bangladesh. https://doi.org/10.56946/jeee.v2i2.513

20. Rahman, J., Foisal, M. Z. U., Mohajan, B., Islam, S., Rafi, A. H., Paul, A., & Ahmad10, S. Role of Renewable Energy, Economic Growth, Agricultural Productivity, and Urbanization Toward Achieving China's Goal of Net-zero Emissions. https://doi.org/10.56946/jeee.v2i1.515

21. Liu, Z., Deng, Z., He, G., Wang, H., Zhang, X., Lin, J., ... & Liang, X. (2022a). Challenges and opportunities for carbon neutrality in China. Nature Reviews Earth & Environment, 3(2), 141-155.

22. Klara, S. M., Srivastava, R. D., & McIlvried, H. G. (2003). Integrated collaborative technology development program for $CO_2$ sequestration in geologic formations—United States Department of Energy R&D. Energy Conversion and Management, 44(17), 2699-2712. https://doi.org/10.1016/S0196-8904(03)00042-6

23. Bleviss, D. L. (2021). Transportation is critical to reducing greenhouse gas emissions in the United States. Wiley Interdisciplinary Reviews: Energy and Environment, 10(2), e390. https://doi.org/10.1002/wene.390

24. Majeed MT, Ayub T (2018) Information and Communication Technology (ICT) and Economic Growth Nexus: A Comparative Global Analysis. Pakistan Journal of Commerce Social Sciences 12(2):443–476

25. Herrmann, M. (2012). Population aging and economic development: anxieties and policy responses. Journal of Population Ageing, 5, 23-46. https://doi.org/10.1007/s12062-011-9053-5

26. Sadorsky P. The effect of urbanization and industrialization on energy use in emerging economies: implications for sustainable development. Am J Econ Sociol. 2014;73(2):392–409. https://doi.org/10.1111/ajes.12072

27. Hu M, Hu XT, Cheng L (2021) Exploring digital economy: a sociosemiotic perspective. Int J Legal Discourse 6:181–202. https://doi.org/10.1515/ijld-2021-2053

28. Murshed M, Nurmakhanova M, Elheddad M, Ahmed R (2020) Value addition in the services sector and its heterogeneous impacts on CO 2 emissions: revisiting the EKC hypothesis for the OPEC using panel spatial estimation techniques. Environ Sci Pollut Res 27(31):38951–38973. https://doi.org/10.1007/s11356-020-09593-4

29. Aslam B, Hu J, Hafeez M, Ma D, AlGarni TS, Saeed M, Abdullah MA, Hussain S (2021) Applying environmental Kuznets curve framework to assess the nexus of industry, globalization, and $CO_2$ emission. Environ Technol Innov 21:101377. https://doi.org/10.1016/j.eti.2021.101377

30. Colvin, R. M., Kemp, L., Talberg, A., De Castella, C., Downie, C., Friel, S., ... & Platow, M. J. (2020). Learning from the climate change debate to avoid polarisation on negative emissions. Environmental Communication, 14(1), 23-35. https://doi.org/10.1080/17524032.2019.1630463

31. Di Vaio, A., Palladino, R., Hassan, R., & Escobar, O. (2020). Artificial intelligence and business models in the sustainable development goals perspective: A systematic literature review. Journal of Business Research, 121, 283-314. https://doi.org/10.1016/j.jbusres.2020.08.019

32. Lammers, T., Rashid, L., Kratzer, J., & Voinov, A. (2022). An analysis of the sustainability goals of digital technology start-ups in Berlin. Technological Forecasting and Social Change, 185, 122096. https://doi.org/10.1016/j.techfore.2022.122096

33. Vinothkumar, J., & Karunamurthy, A. (2022). Recent advancements in artificial intelligence technology: trends and implications. Quing: International Journal of Multidisciplinary Scientific Research and Development, 2(1), 1-11. https://doi.org/10.54368/qijmsrd.2.1.0003

34. He, K., & Hertwich, E. G. (2019). The flow of embodied carbon through the economies of China, the European Union, and the United States. Resources, Conservation and Recycling, 145, 190-198. https://doi.org/10.1016/j.resconrec.2019.02.016

35. Raihan, A., Tanchangya, T., Rahman, J., Ridwan, M., & Ahmad, S. (2022a). The influence of Information and Communication Technologies, Renewable Energies and Urbanization toward Environmental Sustainability in China. Journal of Environmental and Energy Economics, 1(1), 11-23. https://doi.org/10.56946/jeee.v1i1.351

36. Sheraz, M., X. Deyi, J. Ahmed, S. Ullah and A. Ullah (2021). "Moderating the effect of globalization on financial development, energy consumption, human capital, and carbon emissions: evidence from G20 countries." Environ Sci Pollut Res: 1–19. https://doi.org/10.1007/s11356-021-13116-0

37. Koengkan M, Fuinhas JA, Santiago R (2020) The relationship between $CO_2$ emissions, renewable and non-renewable energy consumption, economic growth, and urbanisation in the Southern Common Market. J Environ Econ Policy 9(4):383–401. https://doi.org/10.1080/21606544.2019.1702902

38. Aslam, B., Hu, J., Shahab, S., Ahmad, A., Saleem, M., Shah, S. S. A., ... & Hassan, M. (2021). The nexus of industrialization, GDP per capita and $CO_2$ emission in China. Environmental Technology & Innovation, 23, 101674. https://doi.org/10.1016/j.eti.2021.101674

39. Raihan, A., Atasoy, F. G., Atasoy, M., Ridwan, M., & Paul, A. (2022b). The role of green energy, globalization, urbanization, and economic growth toward environmental sustainability in the United States. Journal of Environmental and Energy Economics, 1(2), 8-17. https://doi.org/10.56946/jeee.v1i2.377

40. Sikder, M., Wang, C., Yao, X., Huai, X., Wu, L., KwameYeboah, F., ... & Dou, X. (2022). The integrated impact of GDP growth, industrialization, energy use, and urbanization on $CO_2$ emissions in developing countries: evidence from the panel ARDL approach. Science of the Total Environment, 837, 155795. https://doi.org/10.1016/j.scitotenv.2022.155795

41. Abbasi, K. R., Adedoyin, F. F., Abbas, J., & Hussain, K. (2021). The impact of energy depletion and renewable energy on $CO_2$ emissions in Thailand: fresh evidence from the novel dynamic ARDL simulation. Renewable Energy, 180, 1439-1450. https://doi.org/10.1016/j.renene.2021.08.078

42. Magazzino, C., Mele, M., & Schneider, N. (2021). A machine learning approach on the relationship among solar and wind energy production, coal consumption, GDP, and $CO_2$ emissions. Renewable Energy, 167, 99-115. https://doi.org/10.1016/j.renene.2020.11.050

43. Solarin, S.A., Nathaniel, S.P., Bekun, F.V. et al. Towards achieving environmental sustainability: environmental quality versus economic growth in a developing economy on ecological footprint via dynamic simulations of ARDL. Environ Sci Pollut Res 28, 17942–17959 (2021). https://doi.org/10.1007/s11356-020-11637-8

44. Mohsin, M., Naseem, S., Sarfraz, M., & Azam, T. (2022). Assessing the effects of fuel energy consumption, foreign direct investment and GDP on $CO_2$ emission: New data science evidence from Europe & Central Asia. Fuel, 314, 123098. https://doi.org/10.1016/j.fuel.2021.123098

45. Masood, A., & Ahmad, K. (2021). A review on emerging artificial intelligence (AI) techniques for air pollution forecasting: Fundamentals, application and performance. Journal of Cleaner Production, 322, 129072. https://doi.org/10.1016/j.jclepro.2021.129072

46. Lilhore, U. K., Simaiya, S., Kaur, A., Prasad, D., Khurana, M., Verma, D. K., & Hassan, A. (2021). Impact of deep learning and machine learning in industry 4.0: Impact of deep learning. In Cyber-Physical, IoT, and Autonomous Systems in Industry 4.0 (pp. 179-197). CRC Press.

47. Ahmed, I., Jeon, G., & Piccialli, F. (2022). From artificial intelligence to explainable artificial intelligence in industry 4.0: a survey on what, how, and where. IEEE Transactions on Industrial Informatics, 18(8), 5031-5042. https://doi.org/10.1109/TII.2022.3146552

48. Kar AK, Choudhary SK, Singh VK (2022) How can artificial intelligence impact sustainability: a systematic literature review. J Clean Prod 376:134120. https://doi.org/10.1016/j.jclepro.2022.134120

49. John N, Wesseling JH, Worrell E, Hekkert M (2022) How key-enabling technologies' regimes influence sociotechnical transitions: the impact of artificial intelligence on decarbonization in the steel industry. J Clean Prod 370:133624. https://doi.org/10.1016/j.jclepro.2022.133624

50. Damioli G, Van Roy V, Vertesy D (2021) The impact of artificial intelligence on labor productivity. Eurasian Bus Rev 11(1):1–25. https://doi.org/10.1007/s40821-020-00172-8

51. Vinuesa R, Azizpour H, Leite I, Balaam M, Dignum V, Domisch S et al (2020) The role of artificial intelligence in achieving the Sustainable Development Goals. Nat Commun 11(1):233. https://doi.org/10.1038/s41467-019-14108-y

51. Dhar P (2020) The carbon impact of artificial intelligence. Nat Mach Intell 2(8):423–425. https://doi.org/10.1038/s42256-020-0219-9

53. Chen, P., Gao, J., Ji, Z., Liang, H., & Peng, Y. (2022). Do artificial intelligence applications affect carbon emission performance?—evidence from panel data analysis of Chinese cities. Energies, 15(15), 5730. https://doi.org/10.3390/en15155730

54. Li Q, Cherian J, Shabbir MS, Sial MS, Li J, Mester I, Badulescu A (2021) Exploring the relationship between renewable energy sources and economic growth. The Case of SAARC Countries. Energies 14(3):520. https://doi.org/10.3390/en14030520

55. Ben Jebli M, Youssef SB, Ozturk I (2016) Testing environmental Kuznets curve hypothesis: the role of renewable and non-renewable energy consumption and trade in OECD countries. Ecol Indic 60:824–831. https://doi.org/10.1016/j.ecolind.2015.08.031

56. Hao, Y., Guo, Y., & Wu, H. (2022). The role of information and communication technology on green total factor energy efficiency: does environmental regulation work?. Business Strategy and the Environment, 31(1), 403-424. https://doi.org/10.1002/bse.2901

57. Baloch, M. A., Mahmood, N., & Zhang, J. W. (2019). Effect of natural resources, renewable energy and economic development on $CO_2$ emissions in BRICS countries. Science of the Total Environment, 678, 632-638. https://doi.org/10.1016/j.scitotenv.2019.05.028

58. Dogan, E., Ozturk, I. The influence of renewable and non-renewable energy consumption and real income on $CO_2$ emissions in the USA: evidence from structural break tests. Environ Sci Pollut Res 24, 10846–10854 (2017). https://doi.org/10.1007/s11356-017-8786-y

59. Salahuddin, M., Habib, M. A., Al-Mulali, U., Ozturk, I., Marshall, M., & Ali, M. I. (2020). Renewable energy and environmental quality: A second-generation panel evidence from the Sub

Saharan Africa (SSA) countries. Environmental research, 191, 110094. https://doi.org/10.1016/j.envres.2020.110094

60. Kartal, M. T., Kılıç Depren, S., Ayhan, F., & Depren, Ö. (2022). Impact of renewable and fossil fuel energy consumption on environmental degradation: evidence from USA by nonlinear approaches. International Journal of Sustainable Development & World Ecology, 29(8), 738-755. https://doi.org/10.1080/13504509.2022.2087115

61. Dagar, V., Khan, M. K., Alvarado, R., Rehman, A., Irfan, M., Adekoya, O. B., & Fahad, S. (2022). Impact of renewable energy consumption, financial development, and natural resources on environmental degradation in OECD countries with dynamic panel data. Environmental Science and Pollution Research, 29(12), 18202-18212. https://doi.org/10.1007/s11356-021-16861-4

62. Apergis, N., & Payne, J. E. (2010). Coal consumption and economic growth: Evidence from a panel of OECD countries. Energy Policy, 38(3), 1353–1359. https://doi.org/10.1016/j.enpol.2009.11.016

63. Farhani, S. (2013). Renewable energy consumption, economic growth and $CO_2$ emissions: Evidence from selected MENA countries. Energy Economics Letters, 1(2), 24–41.

64. Murshed, M., Saboori, B., Madaleno, M., Wang, H., & Doğan, B. (2022). Exploring the nexuses between nuclear energy, renewable energy, and carbon dioxide emissions: the role of economic complexity in the G7 countries. Renewable Energy, 190, 664-674. https://doi.org/10.1016/j.renene.2022.03.121

65. Abbas, S., Gui, P., Chen, A. et al. The effect of renewable energy development, market regulation, and environmental innovation on $CO_2$ emissions in BRICS countries. Environ Sci Pollut Res 29, 59483–59501 (2022). https://doi.org/10.1007/s11356-022-20013-7

66. da Silva, P. P., Cerqueira, P. A., & Ogbe, W. (2018). Determinants of renewable energy growth in Sub-Saharan Africa: Evidence from panel ARDL. Energy, 156, 45–54. https://doi.org/10.1016/j.energy.2018.05.068

67. Huber J (2000) Towards industrial ecology: sustainable development as a concept of ecological modernization. J Environ Pol Plan 2(4):269–285. https://doi.org/10.1080/714038561

68. Wang, J., Dong, K., Dong, X., & Taghizadeh-Hesary, F. (2022). Assessing the digital economy and its carbon-mitigation effects: The case of China. Energy Economics, 113(July), 106198. https://doi.org/10.1016/j.eneco.2022.106198

69. Ma, Q., Tariq, M., Mahmood, H., & Khan, Z. (2022). The nexus between digital economy and carbon dioxide emissions in China: the moderating role of investments in research and development. Technology in Society, 68(November 2021), 101910. https://doi.org/10.1016/j.techsoc.2022.101910

70. Lorek, S., & Spangenberg, J. H. (2014). Sustainable consumption within a sustainable economy–beyond green growth and green economies. Journal of cleaner production, 63, 33-44. https://doi.org/10.1016/j.jclepro.2013.08.045

71. Shobande O A, Ogbeifun L (2021). Has information and communication technology improved environmental quality in the OECD?—a dynamic panel analysis. Int J Sust Dev World 1–11. https://doi.org/10.1080/13504509.2021.1909172

72. Kovacikova, M., Janoskova, P., & Kovacikova, K. (2021). The impact of emissions on the environment within the digital economy. Transportation Research Procedia, 55, 1090–1097. https://doi.org/10.1016/j.trpro.2021.07.080

73. Xu, S., Yang, C., Huang, Z., & Failler, P. (2022). Interaction between digital economy and environmental pollution: New evidence from a spatial perspective. International Journal of Environmental Research and Public Health, 19(9), 5074. https://doi.org/10.3390/ijerph19095074

74. Nguyen, T. T., Pham, T. A. T., & Tram, H. T. X. (2020). Role of information and communication technologies and innovation in driving carbon emissions and economic growth in selected G-20 countries☆. Journal of Environmental Management, 261(January), 110162. https://doi.org/10.1016/j.jenvman.2020.110162

75. Li, Z., & Wang, J. (2022). The dynamic impact of digital economy on carbon emission reduction: evidence city-level empirical data in China. Journal of Cleaner Production, 351(October 2021), 131570. https://doi.org/10.1016/j.jclepro.2022.131570

76. Sumaira, & Siddique, H. M. A. (2023). Industrialization, energy consumption, and environmental pollution: evidence from South Asia. Environmental Science and Pollution Research, 30(2), 4094-4102.

77. Ahmed, F., Ali, I., Kousar, S. et al. The environmental impact of industrialization and foreign direct investment: empirical evidence from Asia-Pacific region. Environ Sci Pollut Res 29, 29778–29792 (2022). https://doi.org/10.1007/s11356-021-17560-w

78. Opoku, E. E. O., & Aluko, O. A. (2021). Heterogeneous effects of industrialization on the environment: Evidence from panel quantile regression. Structural Change and Economic Dynamics, 59, 174-184. https://doi.org/10.1016/j.strueco.2021.08.015

79. Nasir, M. A., Canh, N. P., & Le, T. N. L. (2021). Environmental degradation & role of financialisation, economic development, industrialisation and trade liberalisation. Journal of environmental management, 277, 111471. https://doi.org/10.1016/j.jenvman.2020.111471

80. Mentel, U., Wolanin, E., Eshov, M., & Salahodjaev, R. (2022). Industrialization and $CO_2$ emissions in Sub-Saharan Africa: the mitigating role of renewable electricity. Energies, 15(3), 946. https://doi.org/10.3390/en15030946

81. Kermani FI, Ghasemi M, Abbasi F (2015) Industrialization, electricity consumption and $CO_@$ emissions in Iran. Int J Innov Appl Stud 10(3):969

82. Xu B, Lin B (2015) How industrialization and urbanization process impacts on $CO_2$ emissions in China: evidence from nonparametric additive regression models. Energy Economics 48:188–202. https://doi.org/10.1016/j.eneco.2015.01.005

83. Farooq S, Parveen S, Sahibzada HE (2019) Impact of industrialization, urbanization and energy consumption on environmental degradation: evidence from India. Glob Econ Rev 4(2):1–12

84. Dietz, T., & Rosa, E. A. (1994). Rethinking the environmental impacts of population, affluence and technology. Human ecology review, 1(2), 277-300. https://www.jstor.org/stable/24706840

85. Dietz, T., & Rosa, E. A. (1997). Effects of population and affluence on $CO_2$ emissions. Proceedings of the National Academy of Sciences, 94(1), 175-179. https://doi.org/10.1073/pnas.94.1.175

86. Zhongmin X, Guodong C (2005) The environmental impact of China's population and affluence. J Glaciol Geocryol 5:767–773

87. York R, Rosa EA, Dietz T (2003) STIRPAT, IPAT and ImPACT: analytic tools for unpacking the driving forces of environmental impacts. Ecol Econ. https://doi.org/10.1016/S0921-8009(03)00188-5

88. Arshed, N., Munir, M., & Iqbal, M. (2021). Sustainability assessment using STIRPAT approach to environmental quality: an extended panel data analysis. Environmental Science and Pollution Research, 28(14), 18163-18175. https://doi.org/10.1007/s11356-020-12044-9

89. Caglar, A. E. (2020). The importance of renewable energy consumption and FDI inflows in reducing environmental degradation: bootstrap ARDL bound test in selected 9 countries. Journal of Cleaner Production, 264, 121663. https://doi.org/10.1016/j.jclepro.2020.121663

90. Raihan, A., & Tuspekova, A. (2022). The nexus between economic growth, energy use, urbanization, tourism, and carbon dioxide emissions: New insights from Singapore. Sustainability Analytics and Modeling, 2, 100009. https://doi.org/10.1016/j.samod.2022.100009

91. Dickey, D. A., & Fuller, W. A. (1981). Likelihood ratio statistics for autoregressive time series with a unit root. Econometrica: journal of the Econometric Society, 1057-1072. https://doi.org/10.2307/1912517

92. Elliott, G., Rothenberg, T. J., & Stock, J. H. (1992). Efficient tests for an autoregressive unit root.

93. Phillips, P. C., & Perron, P. (1988). Testing for a unit root in time series regression. biometrika, 75(2), 335-346. https://doi.org/10.1093/biomet/75.2.335

94. Pesaran, M. H., Shin, Y., & Smith, R. J. (2001). Bounds testing approaches to the analysis of level relationships. Journal of applied econometrics, 16(3), 289-326. https://doi.org/10.1002/jae.616

95. McNown, R., Sam, C. Y., & Goh, S. K. (2018). Bootstrapping the autoregressive distributed lag test for cointegration. Applied Economics, 50(13), 1509–1521. https://doi.org/10.1080/00036846.2017.1366643

96. Sam, C. Y., McNown, R., & Goh, S. K. (2019). An augmented autoregressive distributed lag bounds test for cointegration. Economic Modelling, 80, 130–141. https://doi.org/10.1016/j.econmod.2018.11.001

97. Goh, S. K., Yong, J. Y., Lau, C. C., & Tang, T. C. (2017). Bootstrap ARDL on energy-growth relationship for 22 OECD countries. Applied Economics Letters, 24(20), 1464–1467. https://doi.org/10.1080/13504851.2017.1284980

98. Chowdhury, M. I., & Karim, S. R. (2020). Financial Innovations and its Impact on Money Demand in Bangladesh: An Error Correction Model (ECM) Approach. International Review of Business Research Papers, 16(1).

99. Shahid, R., Shijie, L., Yifan, N., & Jian, G. (2022). Pathway to green growth: A panel-ARDL model of environmental upgrading, environmental regulations, and GVC participation for the Chinese manufacturing industry. Frontiers in Environmental Science, 10, 972412. https://doi.org/10.3389/fenvs.2022.972412

100. Hansen, B. E., & Phillips, P. C. (1988). Estimation and inference in models of cointegration: A simulation study.

101. Stock, J. H., & Watson, M. W. (1993). A simple estimator of cointegrating vectors in higher order integrated systems. Econometrica: Journal of the Econometric Society, 61(4), 783–820. https://doi.org/10.2307/2951763

102. Park, J. Y. (1992). Canonical cointegrating regressions. Econometrica: Journal of the Econometric Society, 60(1), 119-143. https://doi.org/10.2307/2951679

103. Begum, R. A., Raihan, A., & Said, M. N. M. (2020). Dynamic impacts of economic growth and forested area on carbon dioxide emissions in Malaysia. Sustainability, 12(22), 9375. https://doi.org/10.3390/su12229375

104. Dogan, E., & Seker, F. (2016). The influence of real output, renewable and non-renewable energy, trade and financial development on carbon emissions in the top renewable energy countries. Renewable and Sustainable Energy Reviews, 60, 1074-1085. https://doi.org/10.1016/j.rser.2016.02.006

105. Granger, C. W. (1969). Investigating causal relations by econometric models and cross-spectral methods. Econometrica: Journal of the Econometric Society, 37(3), 424-438. https://doi.org/10.2307/1912791

106. Winterhalder, M., Schelter, B., Hesse, W., Schwab, K., Leistritz, L., Klan, D., ... & Witte, H. (2005). Comparison of linear signal processing techniques to infer directed interactions in multivariate neural systems. Signal processing, 85(11), 2137-2160. https://doi.org/10.1016/j.sigpro.2005.07.011

107. Afolabi, T. S. (2020). Effect of non-performing loans on microfinance banks' performance in Nigeria: a granger causality approach. https://doi.org/10.9790/487X-2205085763

108. Voumik, L.C., Sultana, T. (2022), Impact of urbanization, industrialization, electrification and renewable energy on the environment in BRICS: Fresh evidence from novel CS-ARDL model. Heliyon, 8(11), e11457. https://doi.org/10.1016/j.heliyon.2022.e11457

109. Majeed, M.T., Khan, S., Tahir, T. (2020), Analytical tool for unpacking the driving forces of environmental impact: An IPAT analysis of Pakistan. Geo Journal, 21, 1-3. https://doi.org/10.1007/s10708-020-10321-1

110. Kirikkaleli, D., Adebayo, T. S., Khan, Z., & Ali, S. (2020). Does globalization matter for ecological footprint in Turkey? Evidence from dual adjustment approach. Environmental Science and Pollution Research 2020 28:11, 28(11), 14009–14017. https://doi.org/10.1007/S11356-020-11654-7

111. Qayyum, M., Ali, M., Nizamani, M. M., Li, S., Yu, Y., Jahanger, A., & Arana, G. (2021). Nexus between financial development, renewable energy consumption, technological innovations and $CO_2$ emissions: The case of India. Energies. https://doi.org/10.3390/en14154505

112. Zubair, A. O., Samad, A. R. A., & Dankumo, A. M. (2020). Does gross domestic income, trade integration, FDI inflows, GDP, and capital reduces $CO_2$ emissions? An empirical evidence from Nigeria. Current Research in Environmental Sustainability, 2, 100009. https://doi.org/10.1016/j.crsust.2020.100009

113. Ali, W., Gohar, R., Chang, B. H., & Wong, W. K. (2022). Revisiting the impacts of globalization, renewable energy consumption, and economic growth on environmental quality in South Asia. Advances in Decision Sciences, 26(3), 1-23.

114. Halliru, A. M., Loganathan, N., Hassan, A. A. G., Mardani, A., & Kamyab, H. (2020). Re-examining the environmental Kuznets curve hypothesis in the Economic Community of West African States: A panel quantile regression approach. Journal of Cleaner Production, 276, 124247. https://doi.org/10.1016/j.jclepro.2020.124247.

115. Cifuentes J, Marulanda G, Bello A, Reneses J (2020) Air temperature forecasting using machine learning techniques: a review. Energies 13(16):4215. https://doi.org/10.3390/en13164215

116. Ridwan WM, Sapitang M, Aziz A, Kushiar KF, Ahmed AN, El-Shafie A (2020) Rainfall forecasting model using machine learning methods: case study Terengganu, Malaysia. Ain Shams Eng J. https://doi.org/10.1016/j.asej.2020.09.011

117. Ham Y-G, Kim J-H, Luo J-J (2019) Deep learning for multi-year ENSO forecasts. Nature 573(7775):568–572. https://doi.org/10.1038/s41586-019-1559-7

118. Chattopadhyay A, Hassanzadeh P, Pasha S (2020) Predicting clustered weather patterns: a test case for applications of convolutional neural networks to spatio-temporal climate data. Sci Rep 10(1):1317. https://doi.org/10.1038/s41598-020-57897-9

119. Awan, A.M., Azam, M. Evaluating the impact of GDP per capita on environmental degradation for G-20 economies: Does N-shaped environmental Kuznets curve exist?. Environ Dev Sustain 24, 11103–11126 (2022). https://doi.org/10.1007/s10668-021-01899-8

120. Sharmin, M. (2021), Relationship of renewable and non-renewable energy utilization with $CO_2$ emission of Bangladesh. Energy Economics Letters, 8(1), 49-59.

121. Waheed, R., Chang, D., Sarwar, S., & Chen, W. (2018). Forest, agriculture, renewable energy, and $CO_2$ emission. Journal of Cleaner Production, 172, 4231-4238. https://doi.org/10.1016/j.jclepro.2017.10.287

122. Sharif A, Mishra S, Sinha A et al (2020) The renewable energy consumption-environmental degradation nexus in Top-10 polluted countries: fresh insights from quantile-on-quantile regression approach. Renew Energy 150:670–690. https://doi.org/10.1016/J.RENENE.2019.12.149

123. Yurtkuran, S. (2021). The effect of agriculture, renewable energy production, and globalization on $CO_2$ emissions in Turkey: A bootstrap ARDL approach. Renewable Energy, 171, 1236-1245. https://doi.org/10.1016/j.renene.2021.03.009

124. Lee, J. W. (2019). Long-run dynamics of renewable energy consumption on carbon emissions and economic growth in the European union. International Journal of Sustainable Development and World Ecology. https://doi.org/10.1080/13504509.2018.1492998

125. Shahbaz, M., Wang, J., Dong, K., & Zhao, J. (2022). The impact of digital economy on energy transition across the globe: The mediating role of government governance. Renewable and Sustainable Energy Reviews, 166(February), 112620. https://doi.org/10.1016/j.rser.2022.112620

126. Song, M., Zheng, C., & Wang, J. (2022). The role of digital economy in China's sustainable development in a post-pandemic environment. Journal of Enterprise Information Management, 35(1), 58-77. https://doi.org/10.1108/JEIM-03-2021-0153

127. Kuntsman, A., & Rattle, I. (2019). Towards a paradigmatic shift in sustainability studies: A systematic review of peer reviewed literature and future agenda setting to consider environmental (Un) sustainability of digital communication. Environmental Communication, 13(5), 567-581. https://doi.org/10.1080/17524032.2019.1596144

128. Guo J, Liang S (2021) The impact mechanism of the digital economy on China's total factor productivity: an uplifting effect or a restraining effect? S China Econ J 10:9–27

129. Ozturk, I., & Ullah, S. (2022). Does digital financial inclusion matter for economic growth and environmental sustainability in OBRI economies? An empirical analysis. Resources, Conservation and Recycling, 185, 106489. https://doi.org/10.1016/j.resconrec.2022.106489

130. Mahmood, H., Alkhateeb, T. T. Y., & Furqan, M. (2020). Industrialization, urbanization and $CO_2$ emissions in Saudi Arabia: Asymmetry analysis. Energy Reports, 6, 1553-1560. https://doi.org/10.1016/j.egyr.2020.06.004

131. Khan, M.K., Teng, J.Z., Khan, M.I., Khan, M.O.: Impact of globalization, economic factors and energy consumption on $CO_2$ emissions in Pakistan. Sci. Total Environ. 688, 424–436 (2019). https://doi.org/10.1016/j.scitotenv.2019a.06.065

132. Zafar A, Ullah S, Majeed MT, Yasmeen R (2020) Environmental pollution in Asian economies: does the industrialisation matter? OPEC Energy Review 44(3):227–248. https://doi.org/10.1111/opec.12181

133. Phong, L. H., Van, D. T. B., & Bao, H. H. G. (2018). The role of globalization on carbon dioxide emission in Vietnam incorporating industrialization, urbanization, gross domestic product per capita and energy use. Int J Energ Econ Policy 8(6), 275–283. https://doi.org/10.32479/ijeep.7065

134. Elfaki, K. E., Khan, Z., Kirikkaleli, D., & Khan, N. (2022). On the nexus between industrialization and carbon emissions: evidence from ASEAN+ 3 economies. Environmental Science and Pollution Research, 1-10. https://doi.org/10.1007/s11356-022-18560-0